\documentclass[prl,twocolumn,superscriptaddress,floatfix]{revtex4-1}
\usepackage{amsmath,amsfonts,amssymb,amsthm,graphics,graphicx,epsfig,bbm}
\usepackage[colorlinks=true,citecolor=blue,linkcolor=blue,urlcolor=blue]{hyperref}
\usepackage[usenames]{color}
\usepackage{graphicx}
\usepackage[caption=false]{subfig}

\usepackage{amsmath}
\usepackage{epsfig}
\usepackage{dcolumn}
\usepackage{bm}
\usepackage{color}
\usepackage{verbatim}
\usepackage{epstopdf}
\usepackage{amssymb} 
\usepackage{amstext}
\usepackage{latexsym}
\usepackage{hyperref}
\usepackage{amsfonts}
\usepackage{psfrag}
\usepackage{xcolor}
\usepackage{times}

\newcommand{\comu}[2]{\left[#1,#2\right]}
\newcommand{\acomu}[2]{\left\{#1,#2\right\}}
\newcommand{\avg}[1]{\left<#1\right>}
\newcommand{\eavg}[1]{\langle\langle#1\rangle\rangle}

\newcommand{\dd}{\mathrm{d}}

\DeclareMathOperator{\Tr}{Tr}
\DeclareMathOperator{\dt}{\frac{\text d}{\text d t}}

\begin{document}
\title{Quantum Memristors}
\author{P. Pfeiffer}
\affiliation{Department of Physical Chemistry, University of the Basque Country UPV/EHU, Apartado 644, E-48080 Bilbao, Spain}
\author{I. L. Egusquiza}
\affiliation{Department of Theoretical Physics and History of Science, University of the Basque Country UPV/EHU, Apartado 644, E-48080 Bilbao, Spain}
\author{M. Di Ventra}
\affiliation{Department of Physics, University of California, San Diego, La Jolla, CA 92093, USA}
\author{M. Sanz} 
\email{mikel.sanz@ehu.eus}
\affiliation{Department of Physical Chemistry, University of the Basque Country UPV/EHU, Apartado 644, E-48080 Bilbao, Spain}
\author{E. Solano}
\affiliation{Department of Physical Chemistry, University of the Basque Country UPV/EHU, Apartado 644, E-48080 Bilbao, Spain}
\affiliation{IKERBASQUE, Basque Foundation for Science, Maria Diaz de Haro 3, 48013 Bilbao, Spain}

\begin{abstract} 
Technology based on memristors, resistors with memory whose resistance depends on the history of the crossing charges, has lately enhanced the classical paradigm of computation with neuromorphic architectures. However, in contrast to the known quantized models of passive circuit elements, such as inductors, capacitors or resistors, the design and realization of a quantum memristor is still missing. Here, we introduce the concept of a quantum memristor as a quantum dissipative device, whose decoherence mechanism is controlled by a continuous-measurement feedback scheme, which accounts for the memory. Indeed, we provide numerical simulations showing that memory effects actually persist in the quantum regime. Our quantization method, specifically designed for superconducting circuits, may be extended to other quantum platforms, allowing for memristor-type constructions in different quantum technologies. The proposed quantum memristor is then a building block for neuromorphic quantum computation and quantum simulations of non-Markovian systems.
\end{abstract}

\date{\today}

\maketitle

Although they are often misused terms, the difference between information storage and memory is relevant. While the former refers to save information in a physical device for a future use without changes, a physical system shows memory when its dynamics, usually named non-Markovian~\cite{BP02,Gardiner2010}, depend on the past states of the system. Recently, there is a growing interest in memristors, resistors with history-dependent resistance which provide memory effects in form of a resistive hysteresis~\cite{Chua1971, Strukov2008}. In this sense, memristors, due to their memory capabilities~\cite{Yang2013, UniversalMemcomputing, TRBdV15}, offer novel applications in information processing architectures. 

A classical memristor is a resistor whose resistance depends on the record of the electrical signals, namely voltage or charges, applied to it. The information about the electrical history is contained in the physical configuration of the memristor, summarized in its internal state variable $\mu$, which enters the (voltage-controlled) memristor $I$-$V$-relationship via~\cite{PhysPropMemr},
%\begin{subequations} \label{eq:clmemristor}
\begin{eqnarray} \label{eq:clmemristor}
I(t)&=&G(\mu(t)) V(t), \label{eq:memrresp}\\
\dot \mu(t) &=& f(\mu(t),V(t)). \label{eq:memrstate}
\end{eqnarray}
%\end{subequations}
The state variable dynamics, encoded in the real-valued function $f(\mu(t), V(t))$ and the state variable-dependent conductance function $G(\mu(t)) > 0$, leads to a characteristic pinched hysteresis loop of a memristor under a periodic driving.

\begin{figure}[h!]  
         \includegraphics[width=0.47\textwidth]{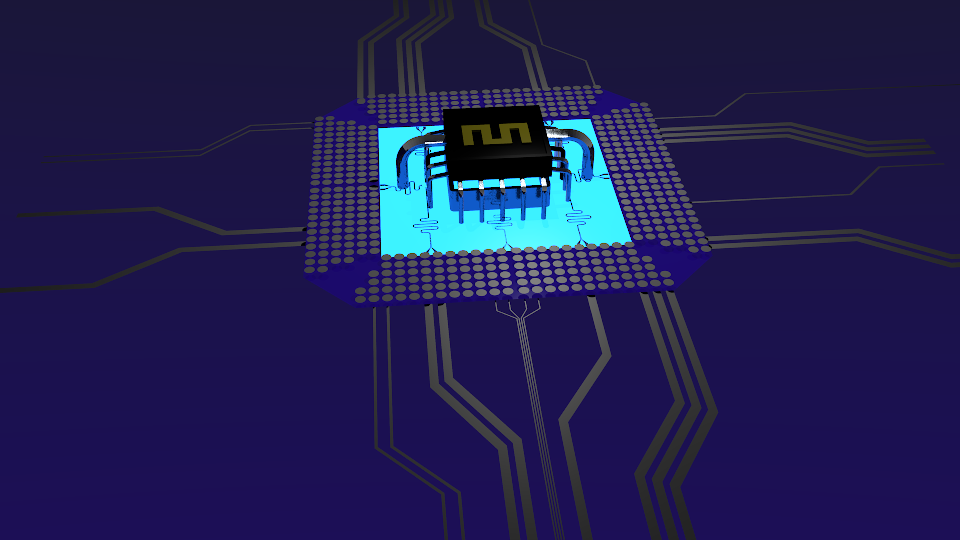}
\caption{Artist view of a quantum memristor coupled to a superconducting circuit in which there is an information flow between the circuit and the memristive environment.}
\label{fig0}
\end{figure}

In superconducting circuits, electrical signals are quantized and can be used to implement quantum simulations~\cite{Salathe15,Barends15}, perform quantum information tasks~\cite{Felicetti14}, or quantum computing~\cite{Barends15-2,Riste15}. However, despite its prospects in classical information processing, memristors have not been considered so far in quantized circuits. The quantum regime of electrical signals is accessed by operating superconducting electric circuits at cryogenic temperatures. Their behavior is well-described by canonical quantization of Lagrange models that reproduce the classical circuit dynamics~\cite{Devoret1995}. Furthermore, the effects of dissipative elements like conventional resistors are studied by coupling the circuit to environmental degrees of freedom represented by a transmission line~\cite{Yurke1983} or a bath of harmonic oscillators~\cite{CaldeiraLegg1983}. Yet, previous studies of memory elements in quantized circuits focussed only on non-dissipative components like memcapacitors and meminductors \cite{5247127,0957-4484-24-25-255201,Shevchenko:2016aa}. This is due to the fact that the history-dependent damping requires dissipative potentials, which cannot be cast in simple system-environment frames~\cite{DiVentra2012}. 

\begin{figure}[!t] 
  \includegraphics[width=\linewidth]{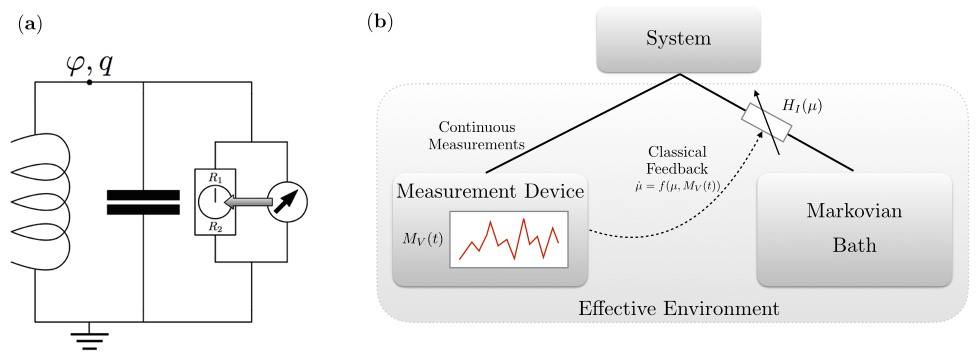}
              \caption{{\bf (a)} Scheme of an LC circuit shunted by a memristor, which following our scheme, it is replaced by a tunable dissipative ohmic environment (depicted by the resistor with a knob to choose a resistance value between $R_1$ and $R_2$), a weak-measurement protocol (depicted by the voltmeter on the right), and a feedback tuning the coupling of the system to the dissipative environment depending on the measurement outcome (represented by the grey arrow). {\bf (b)} Feedback model of a memristor and implementation in quantum dynamics via a feedback-controlled open quantum system.}
              \label{fig1}
\end{figure}

In this Article, we propose a design of a quantum memristor (Fig. \ref{fig0}), closing the gap left by the classical memristor in the quantization toolbox for circuits \cite{PfeifferMSc}. Equation (\ref{eq:memrstate}) can be understood as an information extraction process and, hence, its effect may be modeled on a circuit by continuous measurements. Furthermore, the state-dependent resistance in Eq.~(\ref{eq:memrresp}) is mimicked by a measurement-controlled coupling between the circuit and a bath of harmonic oscillators. Hence, our model for a quantum memristor constitutes a special case of quantum feedback control~\cite{Wiseman2010}, which is not restricted to superconducting circuits. As a paradigmatic example of a quantum memristor, we study a quantum LC circuit shunted by a memristor (see Fig. \ref{fig1}(a)), and address the compatibility between memory effects and quantum properties like coherent superpositions.

\textit{Quantum memristor dynamics} $-$ We decompose the influence of a quantum memristor on the system into a Markovian tunable dissipative environment, a weak-measurement protocol, and a classical feedback controlling the coupling of the system to the dissipative environment. Then, the evolution of the circuit quantum state $\rho$ includes a back action term due to the presence of the weak measurements. In addition, the dynamics contains a dissipative contribution, which depends on the state variable $\mu$, and a Hamiltonian part, so that
\begin{equation}
\dd \rho= \dd \rho_{H}+\dd \rho_{meas}+\dd \rho^{(\mu)}_{damp}, \label{eq:qstatecomp}
\end{equation}
which essentially plays the role of Eq. (\ref{eq:memrresp}). Correspondingly, if the interaction between the state variable and the circuit is cast in a measurement, the voltage records $M_V(t)$ govern the state variable dynamics, 
\begin{equation}
\dot \mu(t) = f(\mu(t),M_V(t)). \label{eq:statewithmeas}
\end{equation}

 \hfill

Finally, the Hamiltonian part is determined by the circuit structure. In the following, we illustrate the method by shunting an LC circuit with a quantum memristor, as depicted in Fig.~\ref{fig1}. Its description requires only one degree of freedom, the top node flux $\varphi$, and its conjugate momentum $q$~\cite{Devoret1995}, which corresponds to the charge on the capacitor connected to the top node. Therefore, we have
 \begin{equation}
 \dd \rho_{H}=-\frac{i}{\hbar}\comu{H(\varphi,q)}{\rho(t)} \dd t \,.\label{eq:rhohami}
 \end{equation}
The measurement of the voltage applied to the quantum memristor implies a monitoring of the node charge $q$, since the voltage determines the charge on the capacitor with capacitance $C$, $q=CV$. Therefore, according to the theory of continuous measurements~\cite{Wiseman2010, Jacobs2006}, the state update and the measurement output have respectively the form
%\begin{subequations}\label{eq:sme}
 \begin{eqnarray}\label{eq:sme}
\dd \rho_{meas}&=&-\frac{\tau}{q_0^2} \comu{q}{\comu{q}{\rho(t)}} \dd t \nonumber\\ &&+ \sqrt{\frac{2\tau}{q_0^2}} \left(\acomu{q}{\rho(t)}-2\avg q \rho(t) \right) \dd W, \label{eq:rhomeas}	\\
M_V(t)&=&\frac{1}{C}\left(\avg{q(t)}+\sqrt{\frac{q_0^2}{8\tau}} \zeta(t)\right),  \label{eq:measout}
 \end{eqnarray}
% \end{subequations}
where $\acomu{A}{B}=AB+BA$ is the anticommutator and the mean value of an observable reads $\avg A= \Tr \left(\rho A\right)$. The projection frequency $\tau$ is defined as the inverse of the measurement time needed to determine the mean charge up to an uncertainty $q_0$, and depends on the measurement strength $k=\frac{\tau}{q_0^2}$. In the limit $\tau\rightarrow \infty$, we recover the usual projective measurement. On the other hand, in the limit $\tau\rightarrow 0$, the measurement apparatus is decoupled from the system, obtaining no information about it. Finally, the stochasticity of the measurement output enters via the white noise $\zeta(t)$, and the Wiener increment $\mathrm d W$ induces the corresponding probabilistic update of the quantum state.

\begin{figure*}
        \subfloat[$\tau=0.005$\label{sfig:t0k005}]{%
  \includegraphics[width=.32\linewidth]{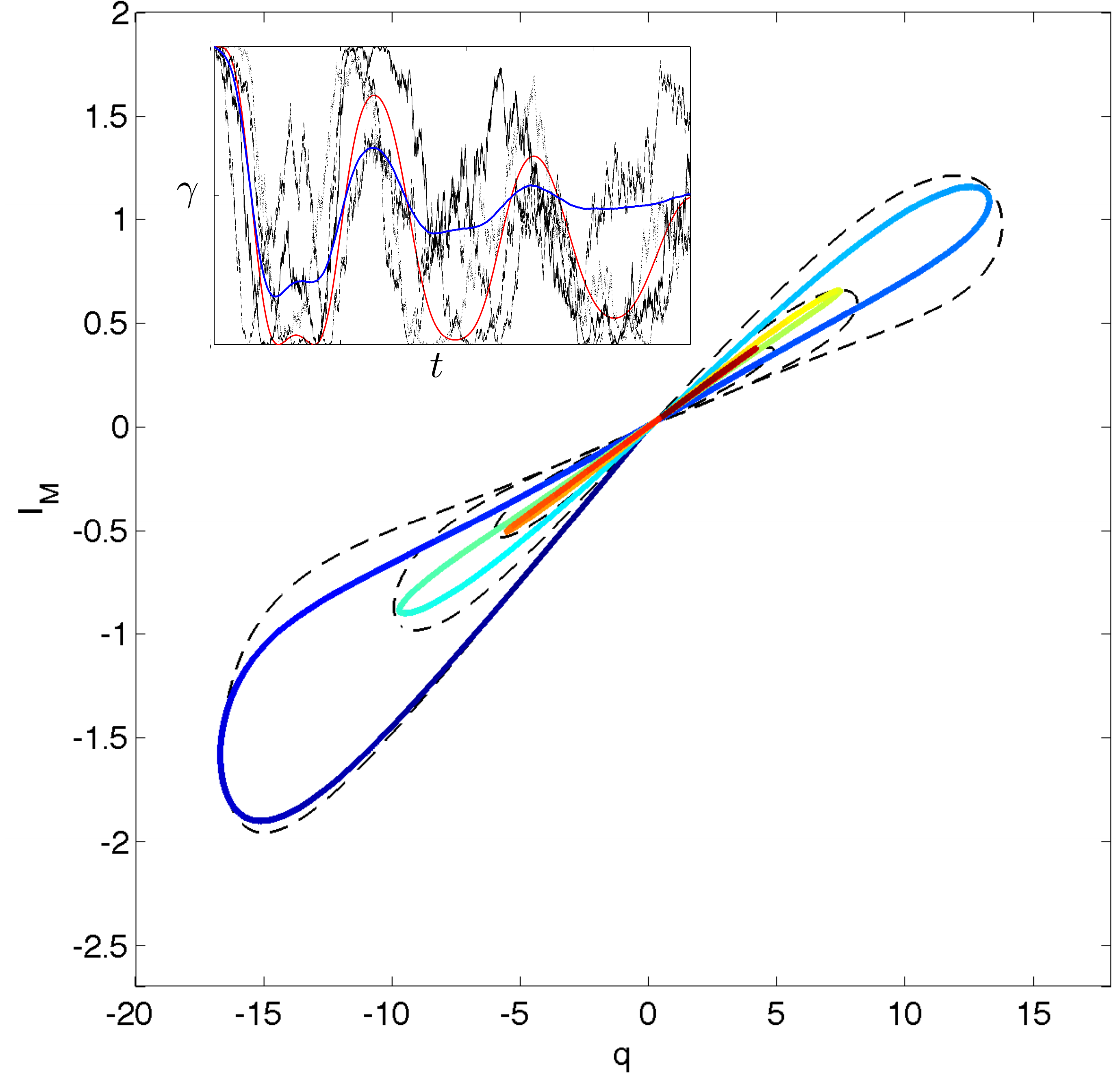}%
}\hfill
\subfloat[$\tau=0.2$\label{sfig:t0k2}]{%
  \includegraphics[width=.32\linewidth]{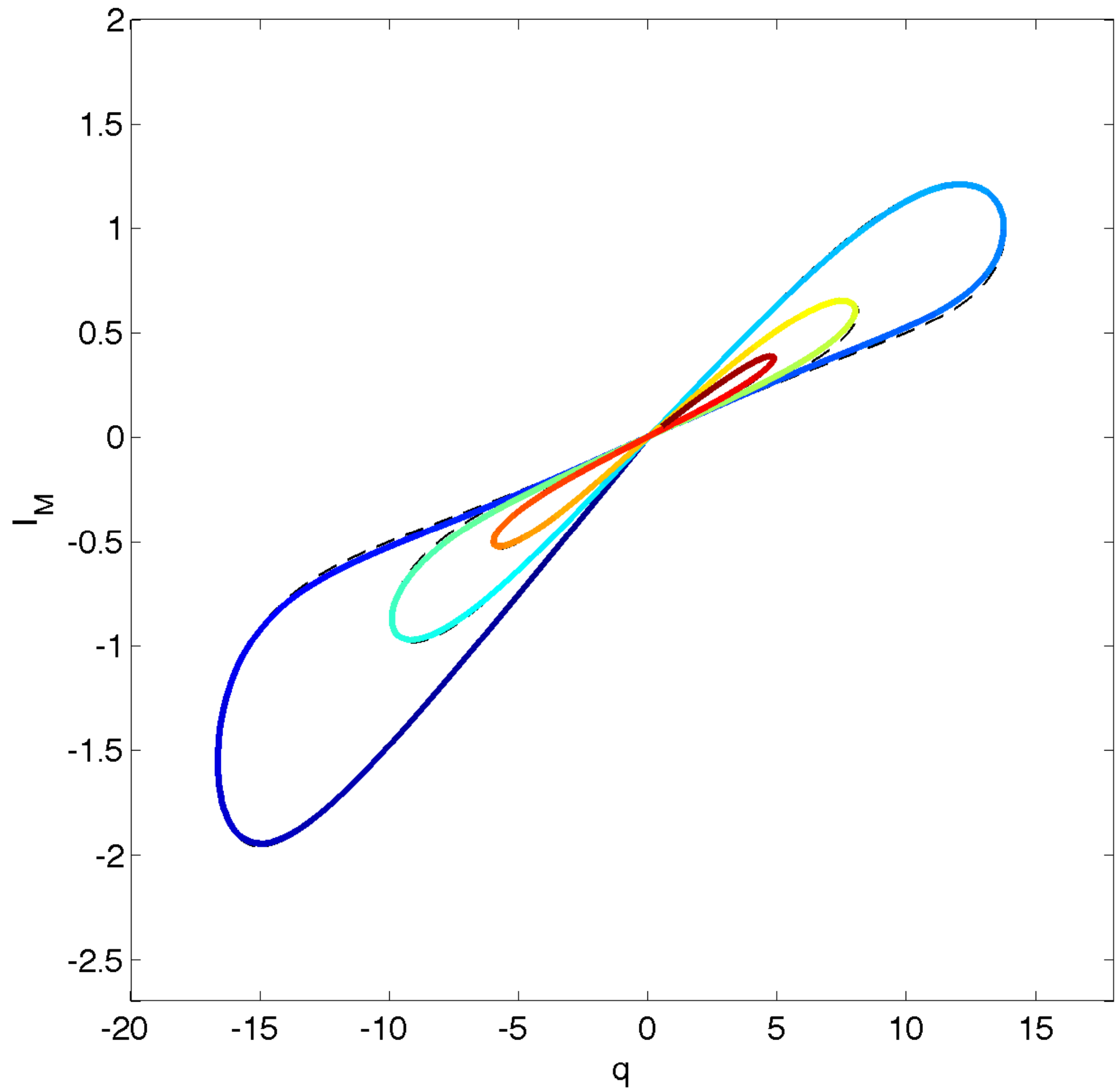}%
}\hfill
\subfloat[$\tau=4$\label{sfig:t4}]{%
  \includegraphics[width=.32\linewidth]{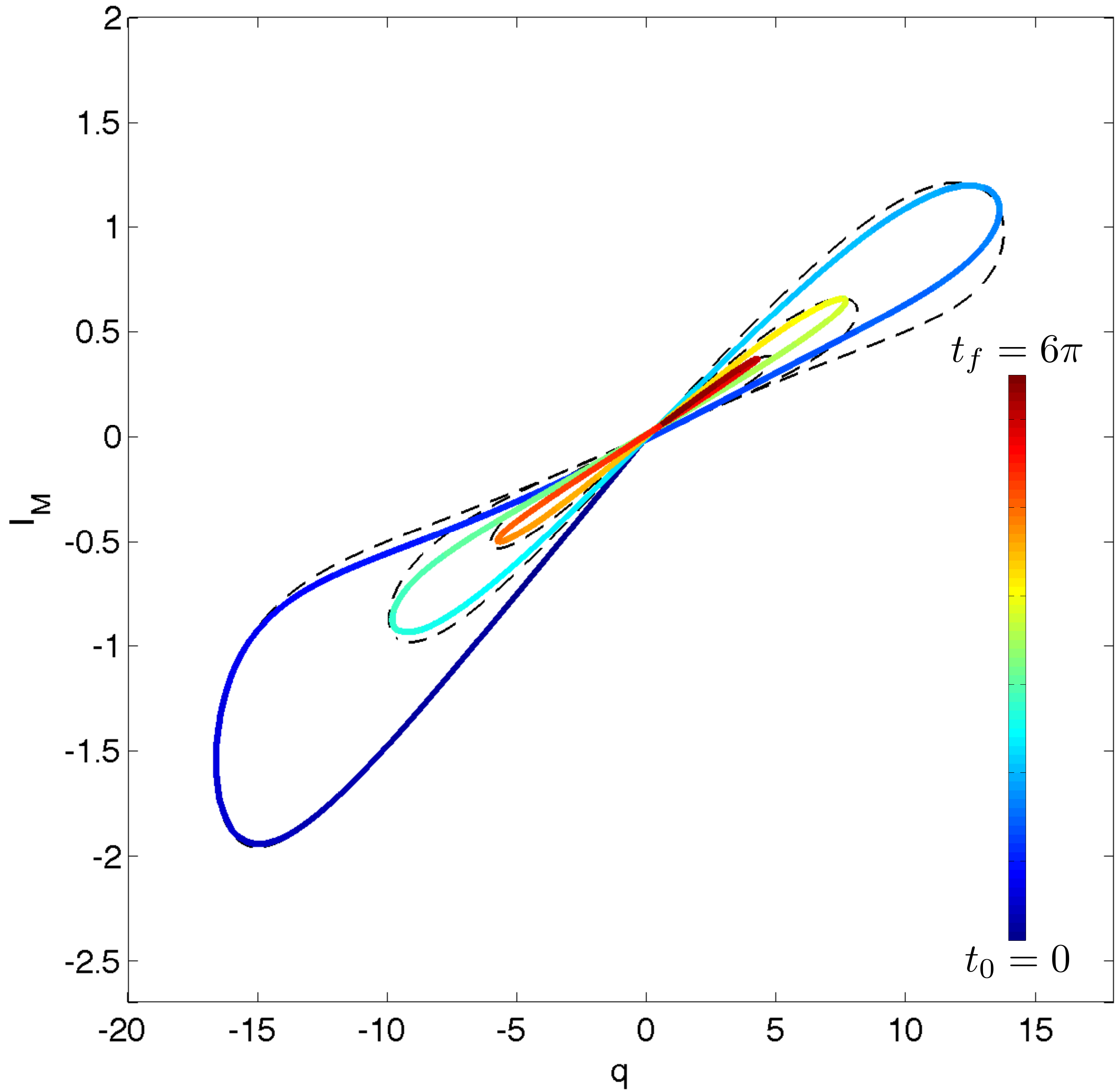}%
}
              \caption{Hysteresis plots of the memristor for the unconditioned evolution with increasing projection frequencies $\tau$. The comparison with the classical hysteresis curve (black, dashed) shows the collapse in the case of a very strong or very weak measurement. The inset shows the evolution of the damping rate and depict several of the underlying stochastic trajectories corresponding to one realisation of the conditioned dynamics. The parameters are $\gamma_0=0.1, \epsilon=0.5, \lambda=10, \nu=0.1$, the initial conditions are $\avg \varphi(0)=20, \avg q(0)=0, V_{\varphi}=0.5, V_q=0.5, C_{\varphi,q}=0, \mu(0)=0$ and the average was obtained by generating 3000 trajectories of the stochastic dynamics via the Euler algorithm with $\dd t =10^{-3}$}
        \label{fig:hyst}
\end{figure*}

A constant resistor with resistance $R$ can be simulated by a bath of LC circuits with an ohmic spectral density~\cite{Devoret1995},
\begin{equation}
J_{ohm}(\omega)=\frac{2C\gamma}{\pi} \omega \frac{\Omega^2}{\Omega^2+\omega^2}.
\end{equation}
Here, the relaxation rate of the circuit is $\gamma=\frac{1}{2RC}$ and $\Omega$ denotes a cut-off frequency. In the high temperature and high cut-off frequency limit, $\lambda:=\frac{k_B T}{\hbar},\Omega \gg \omega_0$, with $\omega_0$ the typical circuit frequency, the dissipative contribution to the circuit dynamics can be cast in the Caldeira-Leggett (C-L) master equation~\cite{CaldeiraLegg1983,BP02}. In a sufficiently small time slice $\mathrm dt$, in which $\mu$ remains approximately constant, a quantum memristor acts like a resistor with resistance $G(\mu)^{-1}$. Therefore,  we adapt the C-L form by replacing the constant relaxation rate $\gamma$ by a function $\gamma(\mu)=\frac{G(\mu)}{2C}$,
\begin{eqnarray}
\dd \rho^{\gamma(\mu)}_{damp}&=& -\frac{i\gamma(\mu)}{\hbar} \comu{\varphi}{\acomu{q}{\rho(t)}}\dd t \nonumber \\ &&- \frac{2C\lambda\gamma(\mu)}{\hbar} \comu{\varphi}{\comu{\varphi}{\rho(t)}}\dd t. \label{eq:rhodamp}
\end{eqnarray}
This phenomenological form could in principle be verified in a two-step procedure. Firstly, by capturing the full joint dynamics of the circuit and the bath when subjected to a feedback control of their interaction Hamiltonian. Secondly, by extracting the circuit dynamics after tracing out the bath degrees of freedom, as depicted in Fig. \ref{fig1}(b). However, the required feedback is non-Markovian, i.e. comprises voltage values from the past, and these systems are generally not analytically tractable~\cite{Wiseman2010}. Still, the C-L form represents a plausible choice, if the ordering $t_{relax} \ll t_{control} \ll t_{exchange} \ll \mathrm d t$ holds for a time-coarse graining (See Supplementary material for a detailed argumentation).

Consider now an observer that has no access to the measurement output \(M_V(t)\). From the observer's point of view, the system evolves with density matrix $\bar \rho = \eavg{\rho}$, where $\eavg \cdot$ denotes the average over all realizations of the Wiener noise. We shall use the term {\it unconditioned state} to refer to \( \bar{ \rho} \). The evolution of the unconditioned state is determined by the ensemble average of Eq.~(\ref{eq:qstatecomp}). Clearly, we do not generally obtain a closed system for \( \bar{ \rho} \) since $\eavg{\gamma(\mu) \rho}$, which appears in the ensemble average of the dissipative term, does not in principle factorize. Now, when it does factorize and \( \eavg{ \gamma(\mu)} \) is constant, the evolution equation of \( \bar{ \rho} \) is of Lindblad form, describing a memoryless quantum Markov process. Thus, it provides us with a witness for non-Markovianity.

To sum up, the quantum-memristor dynamics given by increments of the quantum state in Eqs. (\ref{eq:rhohami}),  (\ref{eq:rhomeas}) and  (\ref{eq:rhodamp}), together with the state variable update in Eq. ~(\ref{eq:statewithmeas}), evolves via two coupled, non-linear stochastic differential equations. Their form is designed to mimic the memory effect due to a memristor, by means of a damping rate which depends on partial information of the history of the quantum state. Unfortunately, this complex dynamics prevents an analytical approach, so we treat it numerically.

\textit{Hysteresis in a quantum memristor} $-$ We present a numerical study of the dynamics of Gaussian states in an LC circuit shunted by a quantum memristor with linear state variable dynamics and the memristance of a Josephson junction~\cite{DiVentra2014SM, Juha2016}. For convenience, charge and flux are expressed in units of their vacuum fluctuations, $\varphi_0=\sqrt{\frac{\hbar}{\omega_0 C}}$ and $q_0=\sqrt{\hbar \omega_0 C}$, with $C$ the capacitance. Furthermore, frequency is expressed in units of the circuit frequency $\omega_0=\frac{1}{\sqrt{LC}}$, where $L$ is the inductance of the coil. Hence, the Hamiltonian reads
\[H=\frac{\hbar}{2} (q^2+\varphi^2).
\] 
This quadratic Hamiltonian, as well as the damping and the measurement contribution to the dynamics, preserve the Gaussianity of an initial state. Therefore, it suffices to follow the evolution of the first and second moments of flux and charge (the charge and flux variances are defined as $V_q=\avg{q^2}-\avg q^2$, $V_\varphi=\avg{\varphi^2}-\avg \varphi^2$ and the covariance reads $C_{\varphi,q}=\frac{1}{2} \avg{\acomu{\varphi}{q}}-\avg \varphi \avg q$). This, together with the damping function $\gamma(\mu)$ and the state variable dynamics, fully determine the dynamics of the conditioned state $\rho$ (see \cite{Jacobs2006}),
%\begin{subequations}\label{eq:gaussstate}
\begin{eqnarray}\label{eq:gaussstate}
\dd  \avg \varphi &=& \avg q \dd t +\sqrt{8 \tau} C_{\varphi,q} \label{eq:avgf} \dd W, \label{eq:eqone}\\
\dd  \avg q &=& - \avg \varphi \dd t - 2\gamma(\mu) \avg q \dd t +\sqrt{8\tau} V_q\dd W, \label{eq:avgq}\\
\dd  V_\varphi &=&2 C_{\varphi,q}\dd t+ 2\tau(1-4 C^2_{\varphi,q}) \dd t, \\
\dd  V_q &=&-2C_{\varphi,q} \dd t-4\gamma(\mu) \left(V_q-\lambda \right) \dd t - 8\tau V^2_q \dd t,\label{eq:varq}\\
\dd  C_{\varphi,q} &=& (V_q-V_\varphi)\dd t - C_{\varphi,q} (2\gamma(\mu)+8\tau V_q) \dd t,\label{eq:covar}\\
\dd \mu &=& \nu\left(\avg{q} \dd t +\frac{\dd W}{\sqrt{8\tau}}\right),\label{eq:stateupdate}\\
\gamma(\mu)&=& \gamma_0(1+\epsilon \cos(\mu)). \label{eq:damprate}
\end{eqnarray}  
%\end{subequations}  
The frequency $\nu$ determines the rate of change of the state variable per unit of charge $q_0$. If the unitless feedback parameter $\epsilon \in [0,1]$ vanishes, the damping rate equals the constant $\gamma_0$, and the system reduces to an LC circuit coupled to a constant resistor and a voltmeter. The specific form chosen for \( \gamma( \mu) \) is inspired in Josephson junction physics (see \cite{Juha2016}), but is only determined here for definiteness.

The set of Eqs.~\eqref{eq:gaussstate}-\eqref{eq:damprate} depends on the projection frequency $\tau$, which is a free parameter, since it is not determined by Eqs.~\eqref{eq:memrresp}-\eqref{eq:memrstate}. Indeed, if we consider the classical limit corresponding to charging the capacitor for $\langle q \rangle \rightarrow \infty$, we recover Eqs.~\eqref{eq:memrresp}-\eqref{eq:memrstate} for every positive $\tau$. Therefore, there is an infinite family of quantum memristors producing the same classical memristive dynamics. However, in the low charging regime, the area of the hysteresis loop in the $I$-$V$-characteristic of the quantum memristor in the unconditioned evolution changes with $\tau$. As the voltage is proportional to the charge on the capacitor and the conductance is proportional to the damping, to obtain the hysteresis loop means to plot $\eavg V \propto \eavg q$ vs. $\eavg{I_M}=\eavg{\gamma(\mu) q}$. From the set of Eqs.~\eqref{eq:gaussstate}-\eqref{eq:damprate}, two sources of diffusion of the state variable $\mu$ can be identified, namely, the noisy measurement output and the stochastic back-action on the first moments (terms in Eqs.~(\ref{eq:avgf}), (\ref{eq:avgq}), and (\ref{eq:stateupdate}) proportional to the Wiener increment $\dd W$). These diffusive terms reduce the hysteresis area, since their physical origins, statistical averaging over multiple voltage histories and respectively, insufficient information extraction, counteract memory effects. Indeed, once the state variable is spread over a range $\geq 2\pi$, the periodicity of the damping rate function leads to a stationary value of its ensemble average, $\eavg \gamma=\gamma_0$, and the hysteresis loop collapses. 

In Fig. \ref{fig:hyst}, the successive collapse of the hysteresis for a strong and a weak measurement case is contrasted with the classical hysteresis, which almost coincides with the hysteresis for the optimal choose of $\tau$, balancing information gain and measurement back-action (cf. Supplementary Material). 

According to our model, the dynamics of the quantum LC circuit has acquired a non-Markovian character by coupling it to a quantum memristor. The question remains, whether characteristics of a genuine quantum system like coherent superpositions are affected by the memristive environment. The existence of non-linear terms in Eqs. \eqref{eq:gaussstate}, which in principle destroy the coherence of the superpositions, makes non-trivial the answer to this question. In any case, one must understand this non-linear behavior as the {\it effective} action of the environment-measurement-feedback protocol, i. e. the quantum memristor, onto the system. This paves the way for employing these quantum memristors as a natural building block for simulating non-linear dynamics or designing non-linear dissipative gates. In particular, and in view of the success of the classical memristor proposal \cite{UniversalMemcomputing,TRBdV15}, this opens the door to a possible development of neuromorphic architectures for quantum computing.

Similarly, one may wonder about the quantumness of the quantum memristor dynamics. Numerically, one can observe oscillations in the squeezing of the quantum state, so that an initial Gaussian state, whose variance is squeezed in momentum $V_q$, periodically changes its squeezing to space, $V_{\varphi}$, during the evolution. In other words, the system density matrix does not commute with itself for different times, which is an evidence of the quantumness of the dynamics \cite{wielandt,2013arXiv1312.1329I,2015arXiv150103099F,2015arXiv151001106J}.

Even though the idea of engineering memristors in the quantum realm seems cumbersome, there are already proposals for employing the memristive component of the Josephson junctions in an asymmetric SQUID in superconducting qubits~\cite{DiVentra2014SM}. Unfortunately, the quantization of that proposed design is not complete, since it is described by a semiclassical model. More recently, a fully quantum realisation of a superconducting quantum memristor has been put forward \cite{Juha2016}. This proposal exploits quasi-particle induced tunneling when supercurrents are cancelled in a Josephson junction, and the parameters explored there are achievable with current technology. This shows, at the very least, that  a quantum memristor will soon be experimentally feasible.

\textit{Conclusion} $-$ We have introduced quantum memristors and presented a protocol to  construct properly the evolution equation for superconducting circuits coupled to quantum memristors. Our model is not restricted to electric circuits and could also be investigated in other quantum platforms like trapped ions or quantum photonics. Besides, we have constructively demonstrated the non-Markovian character of the quantum memristor dynamics, which allowed us to conjecture that the memory effects measured as the area of the hysteresis loop are maximized in the classical limit. Due to the impressive features shown by novel memristor-based computer architectures~\cite{UniversalMemcomputing,TRBdV15,MITIBM}, the quantum memristors proposed here may be considered as a building block for neuromorphic quantum computation and quantum simulation of non-Markovian systems.

\section*{AUTHOR CONTRIBUTIONS}
P.P., as the first author, has been responsible for the development of this work. M.S., with the support of I.L.E., have contributed to the mathematical demonstrations, carried out calculations, and examples. M.S. and E. S. suggested the seminal ideas. M.V. has helped to improve the ideas and results shown in the paper. All authors have carefully proofread the manuscript. E.S. supervised the project throughout all stages.

\section*{ACKNOWLEDGMENTS}
We would like to acknowledge support from Spanish MINECO grant FIS2012-36673-C03-02, Basque Government grants IT-472-10 and IT-559-10, UPV/EHU grant UFI 11/55, PROMISCE and SCALEQIT EU projects. E. S. also acknowledges support from a TUM August-Wilhelm Scheer Visiting Professorship and hospitality of Walther-Meißner-Institut and TUM Institute for Advanced Study. M. D. acknowledges partial support from DOE grant DE-FG02-05ER46204.
 
\section*{ADDITIONAL INFORMATION} 
The authors declare no competing financial interests.

\section*{APPENDIX} 

\subsection{Time scales of the quantum memristor dynamics}

The Caldeira-Leggett (C-L) model describes the dynamics of open quantum systems weakly coupled to a bath of harmonic oscillators at high temperature. Hence, it assumes, first, that the collection of harmonic oscillators is in a thermal state, and, second, that the interaction with the open system is governed by a fixed, weakly interacting Hamiltonian $H_I$ \cite{BP02}.

\begin{figure}[b!]
\centering
\includegraphics[width=0.47 \textwidth]{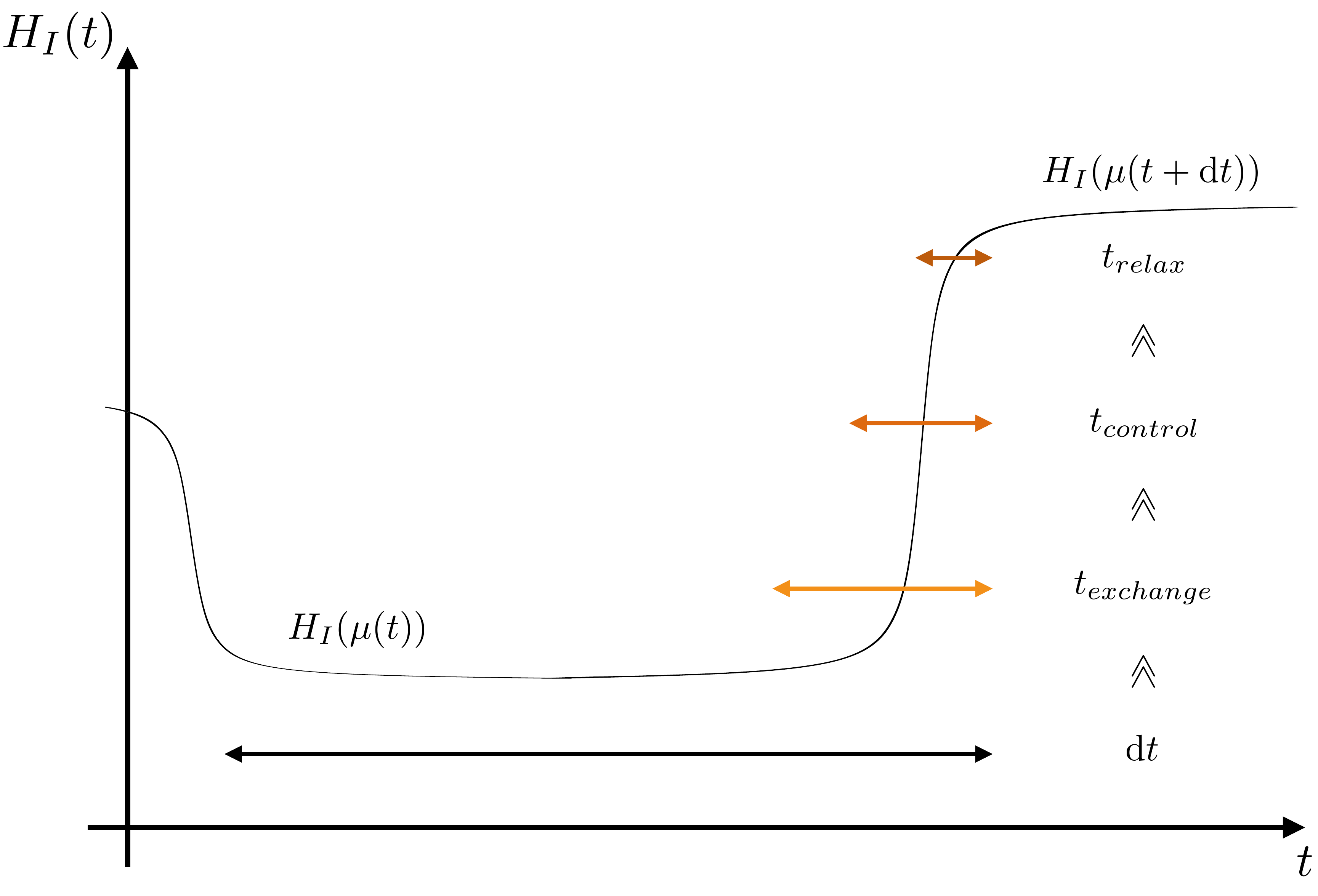}
\caption{Schematic tuning of the interaction Hamiltonian between bath and system. The time ordering of bath relaxation, control time and excitation exchange duration ensures the validity of the Caldeira-Leggett model in every interval $\mathrm d t$ of the coarse graining of time.} 
\label{fig:timescaleordering}
\end{figure}

In the model of the quantum memristor, the interaction Hamiltonian is controlled by  measurement results. The output of measurements is included in  Eq. (5) of the manuscript:
\begin{subequations}\label{eq:sme}
 \begin{eqnarray}
\dd \rho_{meas}&=&-\frac{\tau}{q_0^2} \comu{q}{\comu{q}{\rho(t)}} \dd t \nonumber\\ &&+ \sqrt{\frac{2\tau}{q_0^2}} \left(\acomu{q}{\rho(t)}-2\avg q \rho(t) \right) \dd W, \label{eq:rhomeas}	\\
M_V(t)&=&\frac{1}{C}\left(\avg{q(t)}+\sqrt{\frac{q_0^2}{8\tau}} \zeta(t)\right).  \label{eq:measout}
 \end{eqnarray}
 \end{subequations}
There is a measurement result for every   interval $\mathrm dt$. We assume systematically that the control time $t_{control}$  that is required at each step to fully achieve  the necessary tuning  of the interaction Hamiltonian is  small when compared to the interval $\mathrm dt$. Schematically,  the interaction Hamiltonian does not change  during intervals of length $\mathrm dt$ and does change during intervals of width $t_{control}$, as illustrated in Fig. \ref{fig:timescaleordering}.

We require that during most of the  time interval of length $\mathrm dt$  the bath, and the system interaction with it, be properly described by the C-L model. As a first consequence, we need the thermal structure of the harmonic oscillator bath to be guaranteed. Therefore, the bath relaxation time scale must be shorter than the control time,
\begin{eqnarray*}
t_{relax} \ll t_{control}.
\end{eqnarray*}
Secondly, the interaction between the system and the bath has to be determined by the values of the interaction Hamiltonian in the plateaus, and should not depend on the form of the control step. Specifically,   
the time during which there is exchange of excitations between open system and bath needs to be large compared to the control time,
\begin{eqnarray*}
t_{control} \ll t_{exchange}.
\end{eqnarray*}
In this case, no excitation is exchanged during the tuning of the interaction Hamiltonian. 

In summary, our working assumption is  the time scale ordering shown in Fig. \ref{fig:timescaleordering},
\begin{eqnarray*}
t_{relax} \ll t_{control} \ll t_{exchange} \ll \mathrm d t.
\end{eqnarray*}

\subsection{Classical hysteresis and optimal projection frequency}

\smallskip{{\textit{Classical hysteresis}}

The classical hysteresis curve is obtained by treating the LC circuit coupled to the quantum memristor (see Fig. 2 of the manuscript) classically. The second Kirchhoff's law requires the currents from the three branches, $I_L$ (inductor), $I_C$ (capacitor) and $I_M$ (quantum memristor) to sum to zero, $I_L+I_C+I_M=0$, or equivalently,
\begin{eqnarray}
\frac{\Phi}{L}+\dot Q+G(\mu)\frac{Q}{C} &=& 0. \label{eq:classicalHysteresis1}
\end{eqnarray}
Here, the inductor flux is $\Phi$, the charge on the capacitor is denoted by $Q$, and the memconductance $G$ is a function of the state variable $\mu$. Furthermore, the voltages across the inductor and the capacitor are the same,
\begin{eqnarray}
\dot \Phi = \frac{Q}{C}. \label{eq:classicalHysteresis2}
\end{eqnarray} 
Finally, the second quantum memristor equation (see Eq. (1b) of the manuscript) determines the evolution of the state variable $\mu$, which in our case is a linear function of the applied voltage
\begin{eqnarray}
\dot\mu=\nu \frac{V}{V_0}=\nu \frac{Q}{Q_0}. \label{eq:classicalHysteresis3} 
\end{eqnarray}
Here, the role of the memory frequency $\nu$ becomes clear. It determines the rate of change of the quantum memristor state variable per unit charge $Q_0$ on the capacitor. Clearly, it depends on the choice of this unit charge and we chose the charge fluctuation ($\times \sqrt 2$) $q_0=\sqrt{\hbar\omega_0 C}$ of the LC-circuit in the ground state.

In the dimensionless notation used in the main body of the paper, equations (\ref{eq:classicalHysteresis1}) ,  (\ref{eq:classicalHysteresis2}) and (\ref{eq:classicalHysteresis3}) read 
\begin{eqnarray*} 
\dt q(t) &=& - \varphi(t)-2\gamma(\mu(t)) q(t),\\
 \dt \varphi(t)&=&q(t),\\
 \dt \mu(t)&=&\nu q(t).
 \end{eqnarray*}
 
On solving this set of equations, the  resulting $V$-$I_M$ curve  shows  hysteresis in the voltage-current relation of the memristor for a classical circuit. The black dotted hysteresis curves in the $q$-$\gamma(\mu) q$ plot in Fig. 3 of the manuscript correspond to such solutions, with initial conditions given by the initial expectation values of the corresponding quantum operators $q(0)=\avg q(0)$ and $\varphi(0)=\avg\varphi(0)$, and the same initial state variable value $\mu(0)=\mu_0$.

\subsection{Optimal projection frequency}

\begin{figure}[t!]
\centering
\includegraphics[width=0.47 \textwidth]{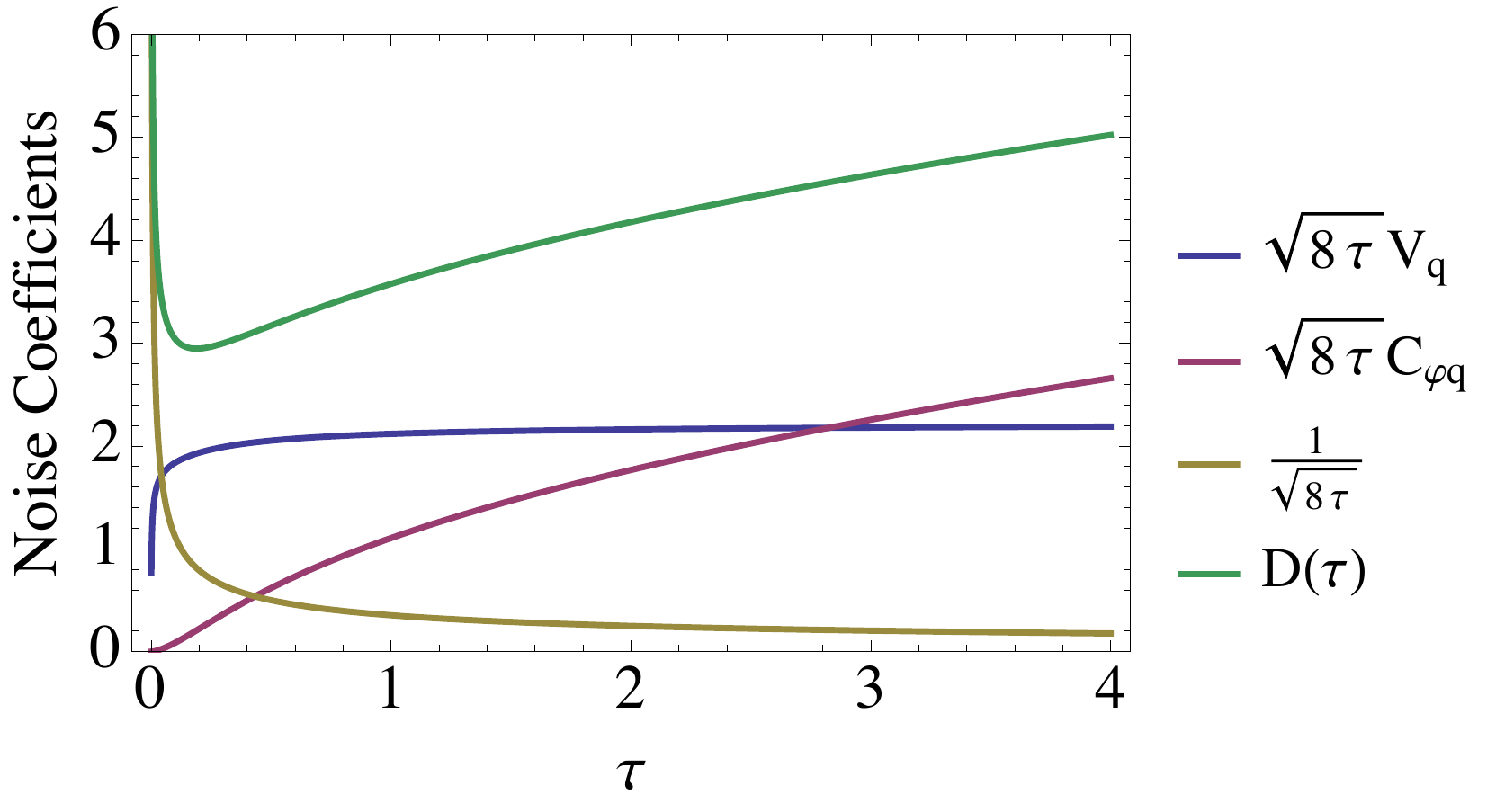}
\caption{ Dependence of all diffusive terms in first moments and  state variable on the projection frequency $\tau$.  The low noise regime suggested by this image is determined by  $\tau\approx 0.2$, for the values of damping rate $\gamma=0.1$ and thermal frequency $\lambda=10$.} 
\label{fig:diffcontrME}
\end{figure}
In our model $\tau$, which is the projection frequency of the measurement, is a free parameter. It measures the rate in which information  is extracted by measurements in the circuit. In a physical implementation of the feedback scheme in the quantum memristor model, it would be tuned by controlling the coupling between the open system and the measurement device. In the classical limit, which corresponds to a large charge in the capacitor, the effective classical memristor equations are independent of $\tau$.

As discussed in the main text, the projection frequency determines  the information gain about the circuit variables as well as their perturbation due to the measurement.  Both uncertainty in the measurement, on the one hand, and backaction, on the other, lead to diffusion of the state variable. Since the projection frequency impacts on both, an optimal \(\tau\) would  correspond to a measuremen, which extracts sufficient information, while inducing only small fluctuations. In other words, the optimal $\tau$ minimises the state variable variance growth.

To determine an approximation of the optimal $\tau$ we analyse the sum of the noise terms in Eqs. (8) in the article,
\begin{equation} 
D(\tau)= \sqrt{8\tau} V_q+ \sqrt{8\tau}|C_{\varphi,q}|+\frac{1}{\sqrt{8\tau}}.
\end{equation} 
The charge variance $V_q$ and the covariance $C_{\varphi,q}$ are a function of time, but they quickly reach quasi stationary values. There are  small variations due to the changes in the damping rate $\gamma$. If we fix the damping rate at its mean value $\gamma_0$ (see Eq. (8g)), the stationary values of Eq. (8d) and (8e) are
\begin{eqnarray}
C_{\varphi,q}^{st}&=&-\frac{1}{8\tau} \left( \sqrt{1+\left(4\tau\right)^2}-1\right),\\
V_q^{st}(\tau)&=&\frac{\sqrt{\gamma_0^2+4\tau(2\gamma_0\lambda-C_{\varphi,q})}-\gamma}{4\tau}.
\end{eqnarray}
Inserting these values in Eqs. (8) for the parameters used in our simulations, the minimisation of the noise sum yields $\tau_{opt}\approx0.2$ (see Fig. \ref{fig:diffcontrME}).

\subsection{Quantum hysteresis is not-pinched at 0}

In the unconditioned evolution of the quantum memristor, the $\eavg q$ vs. $\eavg{\gamma(\mu) q}$ curve is not necessarily pinched at the origin. In fact, numerically evaluating the deviation from factorisation $\delta_q=\eavg{\gamma(\mu) q}-\eavg{\gamma(\mu)}\eavg q$, provided small ($\leq 0.1$), but non-zero values for the zero crossings of $\eavg q$. The resulting hysteresis observed in the unconditioned evolution is depicted in Fig. \ref{fig:nonpinched}. Due to the correlations between damping rate and charge on the capacitor, this corresponds to non-zero (quantum average) current for vanishing (quantum average) voltage. The passivity of the device is still guaranteed, as the expectation value of the emitted power fulfils $\eavg{I_M V}\propto \eavg{\gamma(\mu) q ^2}\geq 0$. 

\begin{figure}[t!]
\centering
\includegraphics[width=0.47\textwidth]{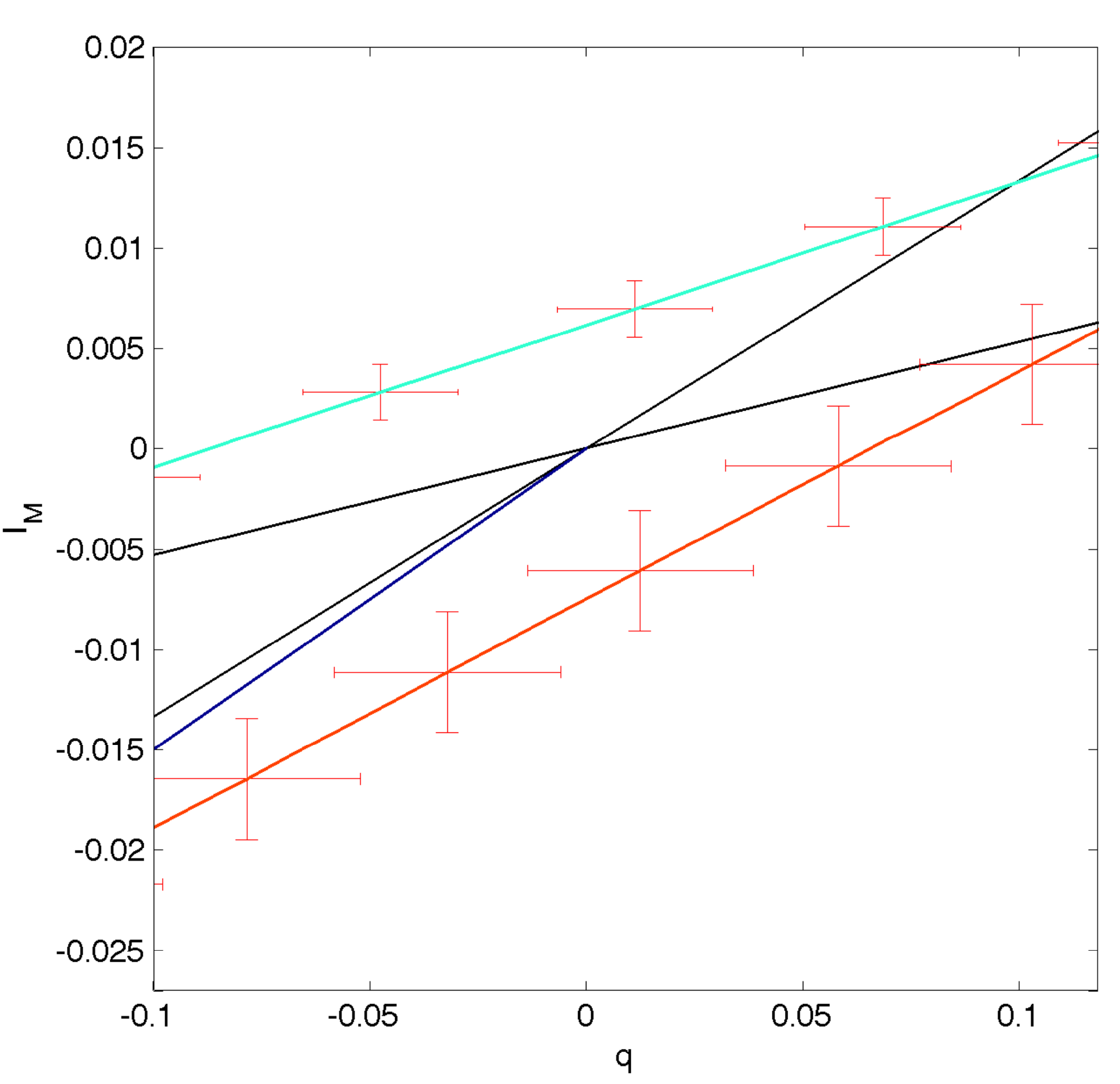}
\caption{Blow-up of the region around the origin of Fig. 3a of the article revealing non pinching at the origin.This is  due to the deviation $\delta_q(t)$ from factorization of $\eavg{\gamma(\mu) q}$. Only  one and a half oscillations are shown, for clarity. The black line corresponds to the classical I-V curve. The red bars denote the ensemble estimation error, not including numerical errors due to the Euler Algorithm.}
\label{fig:nonpinched}
\end{figure}

\subsection{Noisy signals and hysteresis}

Both noisy voltage signals and fluctuations of the state variable have been studied for classical memristors~\cite{PhysPropMemr, Stotland2012}. However, for a classical voltage signal, these two noise sources are independent and can be simultaneously suppressed, whereas in the quantum case of the measurement back-action ties them together. The parameter that controls the noise amplitude is the measurement strength $k$, or more precisely, the projection frequency $\tau = q_0^2 k$ , as defined in the main text. For a strong measurement, (large $\tau\gg1/\omega_0$), noise in the state variable is suppressed, but the back-action increases the fluctuations in the first moments of flux and charge,  thus leading to  diffusion of the state variable. On the contrary, for a weak measurement, $\tau\ll 1/\omega_0$,  first moments evolve in an almost unperturbed manner, but the loss of information in the voltage measurement produces a highly fluctuating state variable update. Finally, the accumulated noise has to be compared to the typical charge on the capacitor, $q_t$, which represents the signal to be recorded, and the threshold $2\pi$, from which on  hysteresis collapses. As a Wiener noise contribution of the form $D \dd W$ induces a linear growth in time of the variance proportional to $D t^2$, the condition for the existence of hysteresis up to a time $t_C$ reads (in adimensionalized terms)
\begin{equation}
8\tau V_q t_C, 8 \tau C_{\varphi,q} t_C, \frac{t_C}{8\tau} \ll \mathrm{min} (q_t^2,4\pi^2). \label{eq:taurestriction}
\end{equation}  
Like all open quantum systems, a quantum memristor is subject to decoherence, a process which has been extensively studied in the context of the quantum-to-classical transition~\cite{Schlosshauer2014}. However, a memory-specific feature of the memristive environment consists in the continuous monitoring of the system with a fine-tuned measurement strength. It is known that continuous measurements are able to provide well-localised trajectories in phase space, a characteristic of classical systems~\cite{Jacobs2006}. In a linear system, such as the LC circuit, the measurement strength has to fulfil two conditions in order for localisation of trajectories to take place, namely, sufficient suppression of the variances and a negligible measurement back action in comparison to the system dynamics. The amplitude of the system dynamics is given by the typical action of the system $s$ (in units of $\hbar$) and constraints the projection frequency by~\cite[Eq. (6)]{Jacobs2000},
\begin{equation} 
\frac{2}{s}\ll \tau \ll 4 s.\label{eq:taurestriction2}
\end{equation}
The typical action can be estimated by $s=\frac{\avg E}{\hbar \omega_0}\approx q_t^2$ (with charge unit \( q_0 \)) and therefore the conditions in Eq. (\ref{eq:taurestriction}) for a memory effect coincide with the requirement for a well-localised trajectory. Loosely speaking, a localised phase space trajectory keeps the state variable localised as well, and thus allows for memory effects.

In order to connect the requirement for a memory effect up to time $t_C$ in Eq.  (\ref{eq:taurestriction}) and the condition for a well-localised phase space trajectory in Eq.  (\ref{eq:taurestriction2}), we first concentrate on fluctuations of first moments. The noise in the first moments induced by the measurement becomes important for large measurement strengths, say $\tau>1/ \omega_0$. In this regime, and using the adimensional setting of the main text, the noise is dominated by the term $\sqrt{8\tau} C_{\varphi,q}$ with $C_{\varphi,q}\approx \frac{1}{2}$ and hence the first two inequalities in  Eq.  (\ref{eq:taurestriction}) require
\begin{equation*}
\frac{8\tau t_C}{2}\ll q_t^2.
\end{equation*}
This corresponds to the upper bound for the projection frequency in Eq.  (\ref{eq:taurestriction2}) growing linearly with the typical action $s$ in units of the planck constant, because  $s=\frac{\avg E}{\hbar\omega}=\frac{q_0^2}{\hbar\omega C} q^2=q^2$.  Correspondingly, the last inequality in Eq.  (\ref{eq:taurestriction})
\begin{equation}
\frac{t_C}{8\tau} \ll q_t ^2 \Rightarrow \frac{t_C}{8 q_t^2}\ll \tau
\end{equation}
provides the inverse scaling with the typical action $s$ of the lower bound on $\tau$ in Eq. (\ref{eq:taurestriction2}).

\subsection{Numeric implementation}

We study  Gaussian state dynamics in an LC circuit coupled to a quantum memristor with numerical simulation of the set of Eqs. (8) in the manuscript. These are stochastic differential equations with Gaussian noise and are cast in  It\^o form. We use the explicit Euler algorithm  \cite{Gardiner2010}. It is based on the discretisation of the equations according to
\begin{eqnarray*}
\mathrm dt &\rightarrow& \Delta t, \\
\mathrm dW &\rightarrow& G \sqrt{\Delta t},
\end{eqnarray*}
where $\Delta t$ has to be chosen sufficiently small and the $G$ are independently, identically, distributed values drawn from a Gaussian distribution with mean 0 and variance 1. 

\begin{figure}[h!]
\centering
\includegraphics[width=0.47\textwidth]{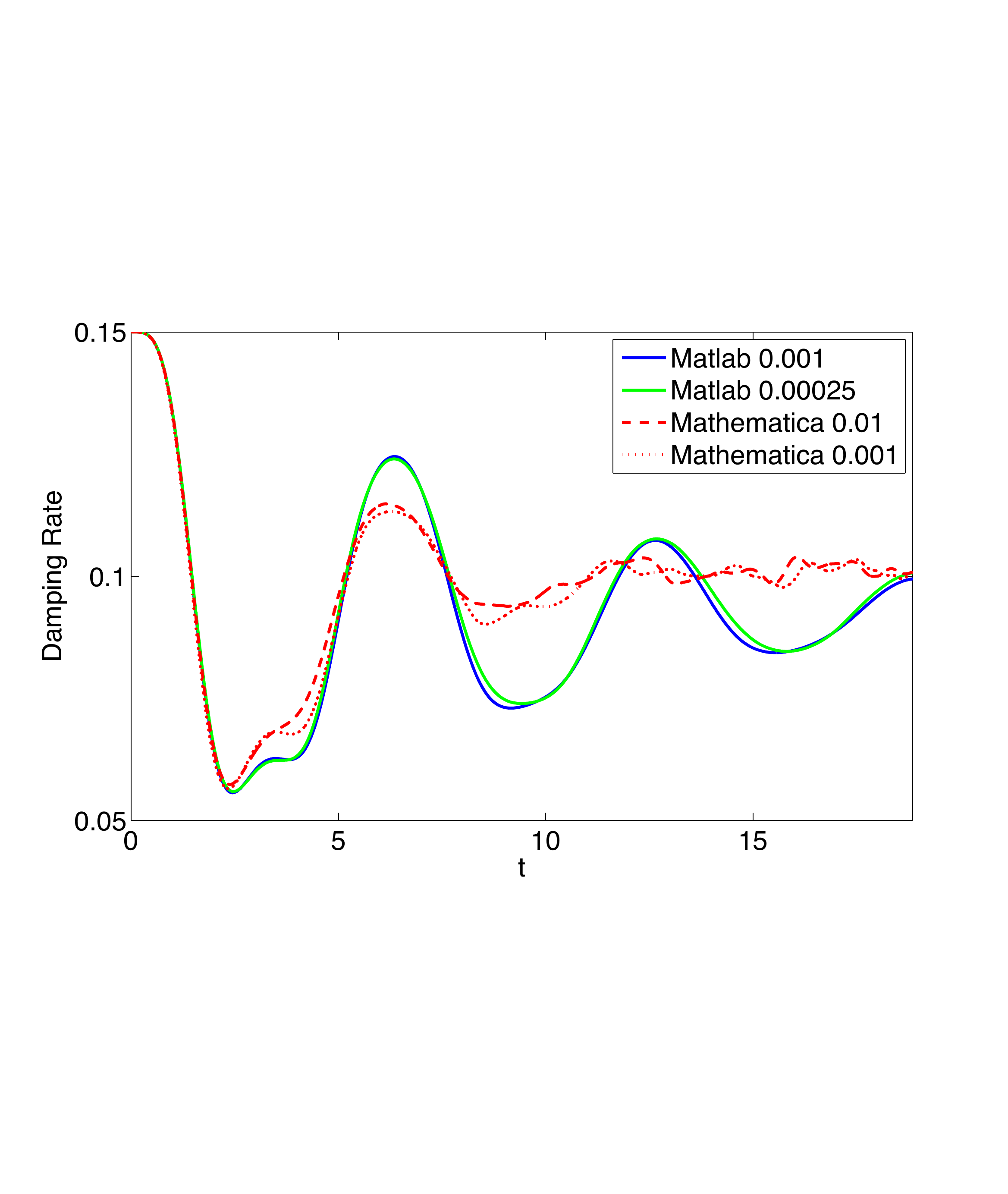}
\caption{Test of the Matlab code via a comparison of the evolution of the mean value damping rate with a solution obtained in Mathematica. The different lines depict different time increments $\dd t$. The ensemble sizes are: 3000 (Matlab), 300 (Mathematica).}
\label{fig:algcomp}
\end{figure}

Concretely, the simulation is implemented in Matlab  (code  available upon request). All the above results are obtained with a time increment $\Delta t = 10^{-3}$, which in the chosen units is much smaller than the period of one oscillation in the LC circuit, $T=2\pi$. In order to obtain the hysteresis curves of the ensemble we average over 3000 trajectories.  

The stability of our implementation with respect to the choice of the time increment and the sufficient ensemble size is demonstrated by a comparison of the evolution of the damping rate $\gamma(\mu(t))$ for different time increments in Fig. \ref{fig:algcomp}. Furthermore, we have also carried out  a simulation using the inbuilt version of the Euler algorithm  in Mathematica, and, qualitatively, the behaviour is the same.  In fact our implementation is more stable for longer times, because of the poor handling of large ensemble sizes in the version of Mathematica available to us.

\end{document}